\documentclass[11pt]{amsart}
\usepackage{geometry}                
\usepackage{graphicx}
\usepackage{amssymb}
\usepackage{epstopdf}
\usepackage{threeparttable}
\title{Application of minimum entropy deconvolution to detect $pP$ phase in a seismogram}
\author{Rong Qiang Wei}
\address{College of Earth and Planetary Sciences, University of Chinese Academy of Sciences, Beijing, PRC, 100049}
\email{wrq1973@ucas.edu.cn}
\date{}

\begin{document}
\maketitle

\begin{abstract}
 The hypocentral depth is a key requirement in seismology and earthquake engineering, but it is very difficult to be determined. The current accepted improvement is taking advantage of the depth phases, such as the pP, to constrain this parameter. However, it is not easy to pick such a phase in a seismogram from the other phases and the backgound noises. Here we propose the use of the minimum entropy deconvolution (MED) to detect it.  Synthetic tests show that impulse(s) hidden in the seimic noises, eg. discrete unit impulses or the Gaussian mono impulses, can be detected completely. Further, we assume that the pP phase is an impulse-like signal buried in the Z component of the seismogram and applied this technique to 12 earthquakes in  the International Association of Seismology and Physics (IASPEI) Ground Truth (GT) reference events list. Results show that 9 out of 12 earthquakes have absolute errors of less than 2.00 s for the travel-time differences of pP-P, and the maximum absolute error is 3.06 s . This demonstrate that the assumption above is reasonable, and this technique works well and effectively even for a single seismogram. Due to its little cost and effectiveness, this technique may be also useful in the starting points for other methods to detect pP phase.      
 
\end{abstract}

{\hspace{2.2em}\small Keywords:}

{\hspace{2.2em}\tiny earthquake, hypocentral depth, pP phase, MEC, Minimum entropy deconvolution (MED)}

\section{Introduction}

The hypocentral depth is one of the four parameters for locating an earthquake. It's accurate determination is important for the study of many fields of the earth science, including continental dynamics, seismic hazard assessment, lithospheric rheology and structure, and plate tectonics. For example, Maggi et al. (2000) suggested that earthquakes in the continents are restricted in the upper brittle part of the crust and are either rare or absent in the underlying mantle. Based on this result, Jackson (2002) proposed a rheological model for the continental lithosphere in which the upper crust is strong but the mantle is weak. On the contrary, Chen and Yang (2004) argued that the mantle lithosphere is sufficiently strong with their observation that 11 earthquakes occurred in the mantle beneath the western Himalayan syntaxis, the western Kunlun Mountains, and southern Tibet. This controversy continues and new models have emerged base on the distribution of the hypocentral depthes (eg., Burov, 2011; Chen et al., 2013; Prieto et al., 2017; Sunilkumar et al., 2019; Schulte-Pelkum et al., 2019).

There is agreement that hypocentral depth is the most difficult parameter to be determined in locating an earthquake, because of the trade-off with the origin time and as it might be biased by lateral earth heterogeneities (Letort et al., 2014). A reliable way to estimate accurately it is through the detection of the depth phases (for example pP, sS, or sP) at teleseismic distances (eg.,Abe, 1974; Engdahl et al., 1998; Bond$\acute{a}$r et al., 2004b; Craig, 2019). The reason is that the travel-time differences of pP-P and/or others (eg., sP-P, sS-S) are quite constant over large range of epicentral distances for a given depth ( This can be found in the IASP91 tables (Kennett and Engdahl, 1991)). Therefore the hypocentral depth can be determined nearly independently of the epicenter distance, although there is indeed a little moveout of the depth phases with increasing epicentral distance (eg., Craig, 2019). For example,  ISC‐EHB have included 569,251 pP phases to constrain its hypocentral depths and the hypocentral depth uncertainties were significantly reduced (Engdahl et al., 2020). 

Various methods have been developed to detect the pP phase, especially the travel-time differences of pP-P. These methods can be roughly classified into two categories depending on stacking or not.  The common methods are (1) The cepstral method in frequency domain (Cohen, 1970; Bonner et al., 2002; Letort et al., 2014, 2015); (2) The time-domain equivalents of the 'cepstral' method (Kemerait and Sutton, 1982; Fang and van der Hilst, 2019); (3) The stacking method (Florez and Prieto, 2017; Craig, 2019). The cepstral method estimates the  travel-time differences of pP-P through the peak 'cepstrum' in amplitude associated with separation times between various seismic phases from the Fourier transform of the (log) frequency spectrum of the waveform. The later two methods are those array-based and take advantage of the fact that the coherent arrivals (energy) can be enhanced while the incoherent noise (energy) can be suppressed by appropriate stacking of the signals across all array stations. With the increasing array networks, the methods based on array will be applied more and more widely. However, in many cases it is still necessary to develop new techniques to detect pP phase, especially for those seismograms observed only on a single-station.    

Here we introduce such a technique of Minimum Entropy Deconvolution (MED).  In the following sections, we will first introduce briefly the minimum entropy criterion (MEC) and MED,  and validate the technique of MED with some synthetic tests. We then apply this technique to 12 earthquakes in IASPEI  GT reference events list  in which the arrivals of pP phases are picked with an high accuracy. Finally, we analyze the result and discuss another approach based on autocorrelation and MEC to detect echoes, and some related problems should to be noted.

\begin{figure}
\includegraphics[scale=0.4]{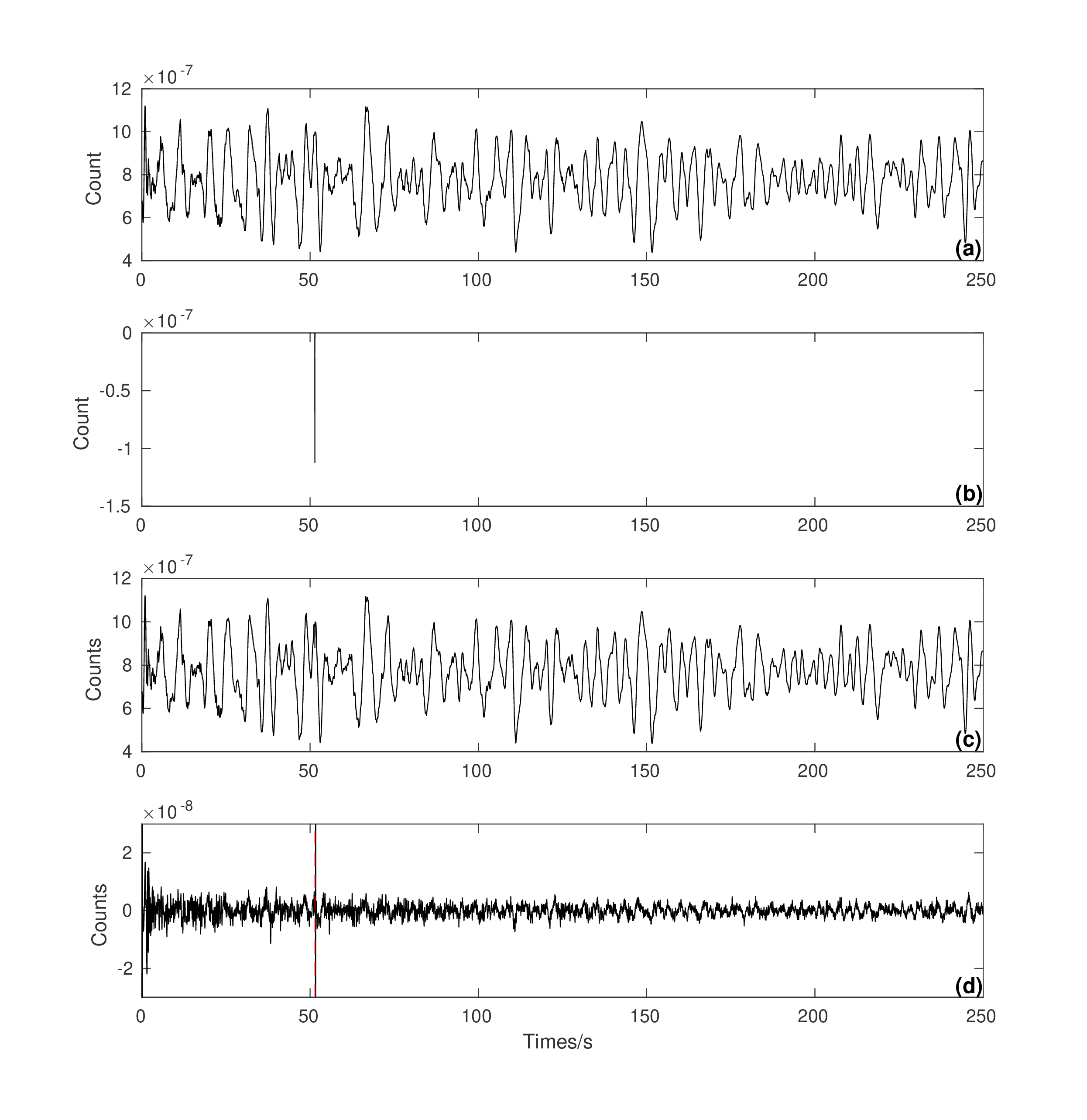} 
\caption{{\footnotesize (a) is the seismogram for the Event  610471076 (Table 1) in IASPEI  GT reference events list, which is recorded at the station DBIC. (b) shows a discrete unit impulse to be added. (c) is the synthetic waveform including the discrete unit impulse. (d) The series of reflectivity obtained by MED (Eq. (\ref{eq2})). The red dashed line marks the time at which the discrete unit impulse is added.} }\label{fig1}
\end{figure}

\section{MED and assumptions}

MED is a technique developed by Wiggins (1978) to extract the reflectivity information from seismic reflection recordings.  In this technique, the seismic recording is assumed as a convolution model of a seismic wavelet with a series of reflectivity impulse response of the earth,

\begin{equation}\label{eq1}
\bf{x}(t)=\bf{w}(t)*\bf{r}(t)
\end{equation}
where $\{x(t);t=1,N\}$ is the sample seismic recording (seismogram), $\{r(t); t=1,N\}$ the reflectivity impulses, and $\{w(t);t=1,M\}$ the seismic wavelet. $*$ denotes the convolution operator, ie., $x(t)=\sum^M_{t'=1}w(t')r(t-t')$.  

Once the deconvolution filter $\bf{\bar{w}}(t)$ ($\bf{\bar{w}}(t)\approx w^{-1}(t)$) is computed, $\bf{\bar{r}}(t)$ (an estimate of $\bf{r}(t)$) is equal to $\bf{\bar{w}}(t)*\bf{x}(t)$, that is,

\begin{equation}\label{eq2}
\bar{r}(t)=\sum^M_{t'=1}\bar{w}(t')x(t-t')
\end{equation}

Computing $\bf{\bar{w}}(t)$ is obviously a blind problem, because only $\bf{x}(t)$ is known in Eq. (\ref{eq1}). 

With the MED technique, $\bf{\bar{w}}(t)$ can be computed by maximizing the Kurtosis norm of the $\bf{\bar{r}}(t)$, ie.,

\begin{equation}\label{eq3}
V_{\bar{r}} =\frac{\sum^N_{_{t=1}}\bar{r}^4(t)}{\left[\sum^N_{_{t=1}}\bar{r}^2(t)\right]^2}
\end{equation}

And $\bf{\bar{w}(t)}$ can be obtained from the Eq. (\ref{eq4})

\begin{equation}\label{eq4}
\bf{R}\bf{\bar{w}}=\bf{g}
\end{equation}
where $R_{it'}=\sum^N_{t=1}x(t-i)x(t-t')$, and $g_i=\frac{\left[\sum_{t=1}^N\bar{r}^3(t)x(t-i)\right]}{V_{\bar{r}}\sum^N_{t=1}\bar{r}^2(t)}$, $i=1,2,..M$.

Eq. (\ref{eq4}) is highly nonlinear so that it can be solved iteratively. That is, assuming a value of $\bf{\bar{w}}$, computing $\bf{R}$ and $\bf{g}$, solving Eq. (\ref{eq4}) for $\bf{g}$, recomputing $\bf{R}$ and $\bf{g}$, etc. Termination is defined as either a number of iterations, or a minimum change in Kurtosis norm between iterations. When $\bf{\bar{w}}(t)$ is obtained, $\bf{\bar{r}}(t)$ can be estimated from Eq. (\ref{eq2}).

The advantage of this MED technique, as compared with other methods, is that it obviates the strong hypotheses
over the phase characteristics of the seismic wavelet and the reflection series. The MED require only the simplicity of the output signal, in which the degree of simplicity is measured with the the Kurtosis norm of the output signal (Eq. (\ref{eq3})). Kurtosis norm has a feature that it is large for impulse-like signals. Wiggins (1978) has applied the MED to exploration seismic recordings and teleseismic core phases (the $PKP$ and $PKKP$). Besides reconstructing the impulse-like signals, Wiggins (1978) found that the MED can reduce the low-frequency noise and is helpful to the phase shift and time separation between overlapping arrivals. Even more, some work demonstrated MED's effectiveness in the fault detection of rotating machine (Endo and Randall, 2007; McDonald and Zhao, 2017).  We will further vindicate the effectiveness of the MED with some synthetic examples.

\begin{figure}
\includegraphics[scale=0.4]{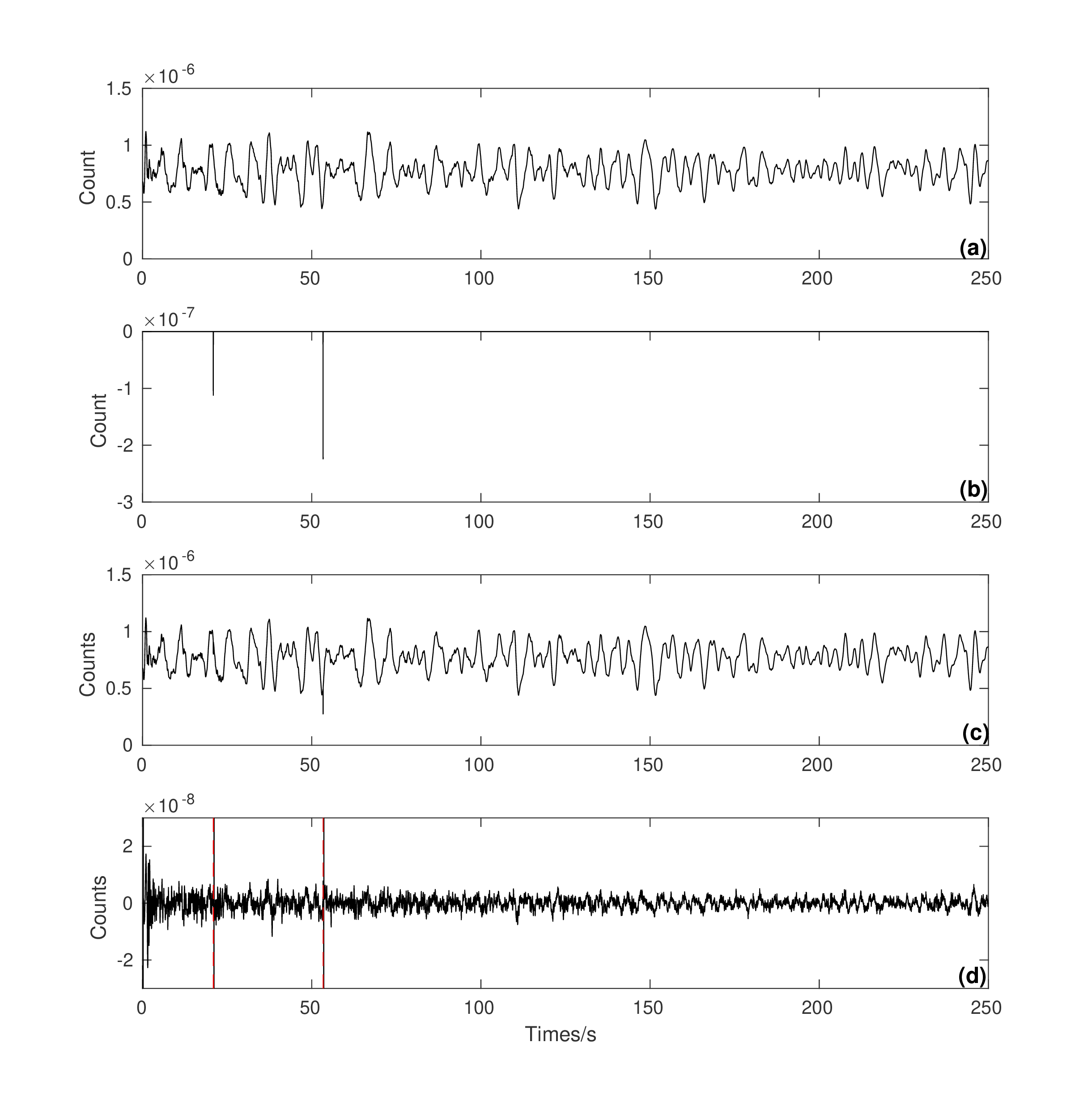} 
\caption{{\footnotesize (a) is the seismogram for the Event  610471076 (Table 1) in IASPEI  GT reference events list, which is recorded at the station DBIC. (b) shows two discrete unit impulses to be added. (c) is the synthetic waveform including the discrete unit impulses. (d) The series of reflectivity obtained by MED (Eq. (\ref{eq2})). The red dashed lines mark the true times at which the discrete unit impulses occur.} }\label{fig2}
\end{figure}

\section{Synthetic Tests}

Our synthetic background data is from a true seismogram for the Event  610471076 (Table 1) in IASPEI  GT reference events list \footnote{International Seismological Centre (2019), IASPEI Reference Event (GT) List, https://doi.org/10.31905/32NSJF7V}(Bond$\acute{a}$r et al., 2004a, 2008; Bond$\acute{a}$r and McLaughlin, 2009). Here and after we only use BHZ component of the seismogram. This seismogram is recorded at the station DBIC.  We cut it using between 0.25s and 250s after the P-wave arrival, which excludes the pP phase  (0.18s after P arrival from the IASPEI  GT bulletin) for this earthquake. Then we add one or two impulses into this seismograms and use MED to detect them.  These impulses include the discrete unit impulse (Dirac) and the Gaussian mono-impulse. Here only the discrete unit impulse with a negative amplitude and the Gaussian mono-impulse with a negative starting amplitude are taken into account, because the pP phase is a solid-surface reflection. 

\begin{figure}
\includegraphics[scale=0.4]{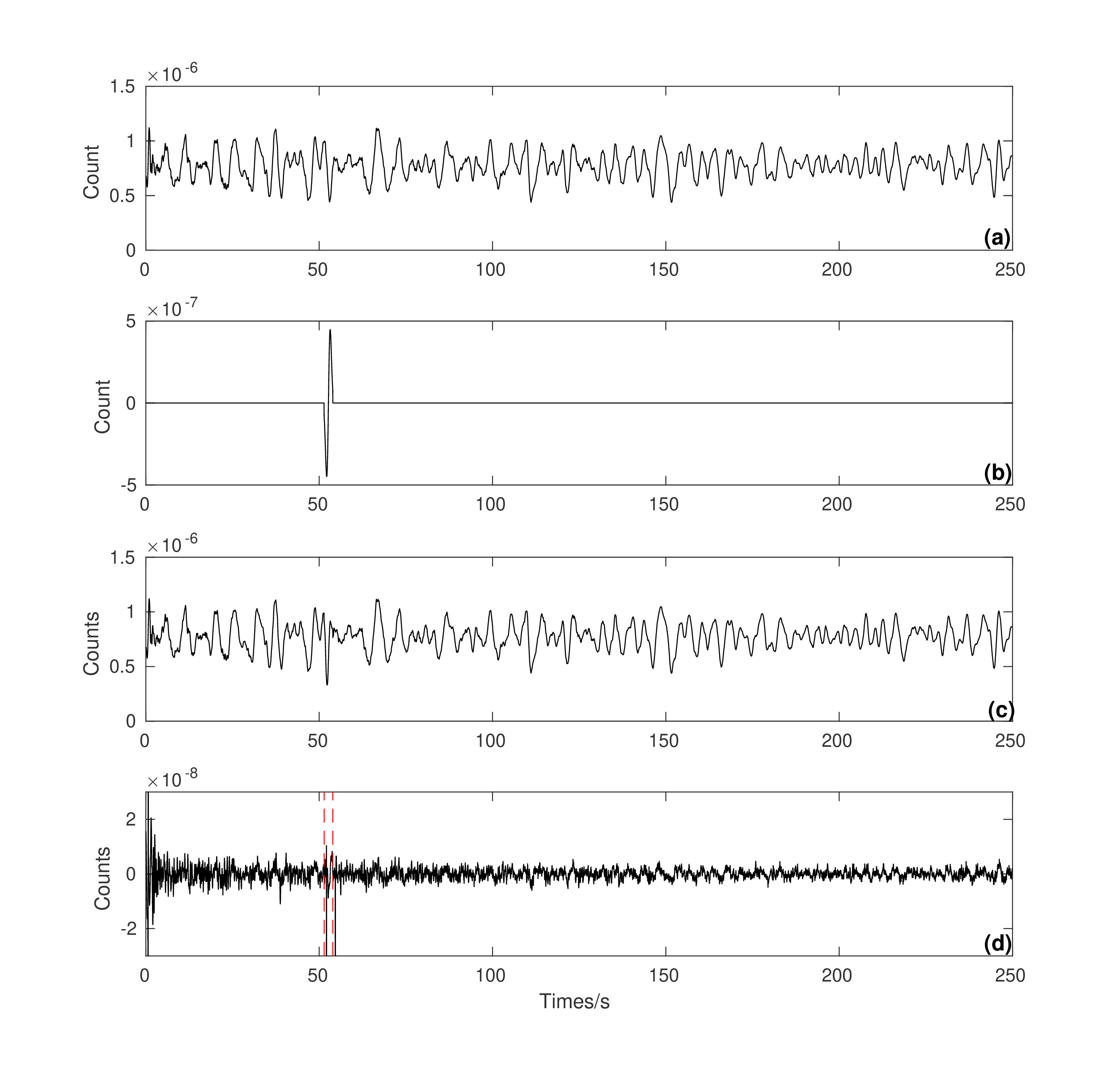} 
\caption{{\footnotesize (a) is the seismogram for the Event  610471076 (Table 1) in IASPEI  GT reference events list, which is recorded at the station DBIC. (b) is a Gaussian mono-impulse to be added. (c) is the synthetic waveform including the Gaussian mono-impulse. (d) The series of reflectivity obtained by MED (Eq. (\ref{eq2})). The red dashed lines mark the true times at which the impulse starts and ends.} }\label{fig3}
\end{figure}

\begin{figure}
\includegraphics[scale=0.4]{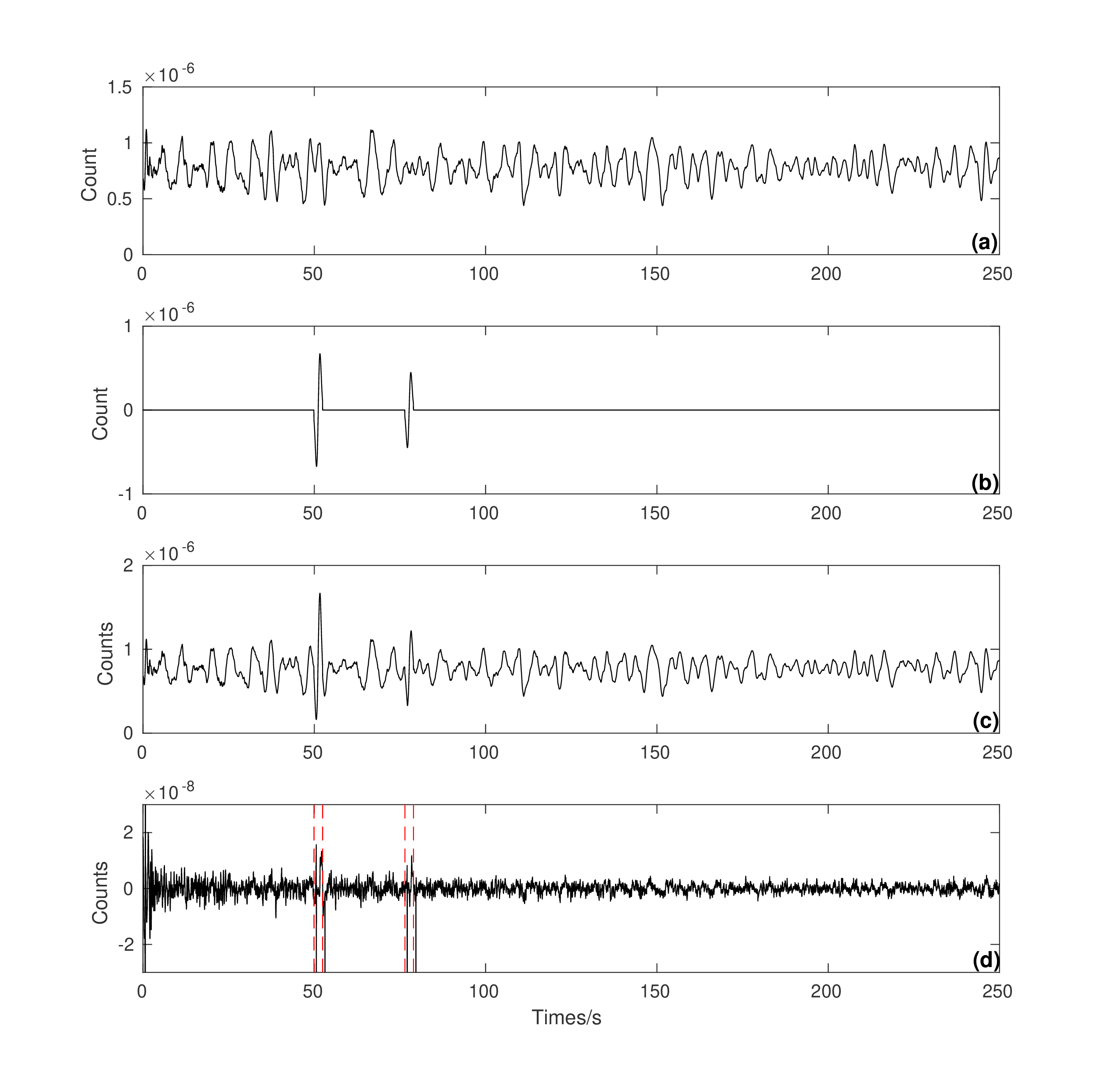} 
\caption{{\footnotesize (a) is the seismogram for the Event  610471076 (Table 1) in IASPEI  GT reference events list, which is recorded at the station DBIC. (b) shows two Gaussian mono-impulses to be added. (c) is the synthetic waveform including the Gaussian mono-impulses. (d) The series of reflectivity obtained by MED (Eq. (\ref{eq2})). The red dashed lines mark the true times at which two impulses start and end.} }\label{fig4}
\end{figure}

Figure {\ref{fig1}}. (a) shows the seismogram we cut; And Figure {\ref{fig1}}. (b) gives a discrete unit impulse to be added, whose amplitude is $-0.1A_{\rm{max}}$ ($A_{\rm{max}}$ is the maximum absolute amplitude of waveform in our time window).  The synthetic data is shown in Figure {\ref{fig1}}. (c), ie., "(a)+(b)". The series of reflectivity (ie., $\bf{\bar{r}}(t)$) obtained by MED (Eq. (\ref{eq2})) are in Figure {\ref{fig1}}. (d), and we mark the true time at which the discrete unit impulse occurs with a red dashed line in this subfigure for comparison. It can be found that the this dashed line is coincident with the line occupying the smallest amplitude. That is to say, the discrete unit impulse buried in the seismogram can be picked up; It is at the time at which the amplitude is the smallest (or the absolute value of the amplitude is the largest) in the series of reflectivity obtained by MED.

Figure {\ref{fig2}} shows the result from burying two discrete unit impulses into the seismogram we used.  These two discrete unit impulses have the amplitude of $-0.1A_{\rm{max}}$ and $-0.2A_{\rm{max}}$, respectively. The series of reflectivity are in Figure {\ref{fig2}}. (d), and we also mark the true times at which the discrete unit impulses occur with red dashed lines in this subfigure. It can be found that the these dashed lines are coincident with the lines occupying the smallest amplitudes. Further analysis shows that the second impulse with the larger amplitude is at the time at which the amplitude is the smallest, while the first impulse is located at the time at which the amplitude is the second smallest.

\begin{figure}
\includegraphics[scale=0.5]{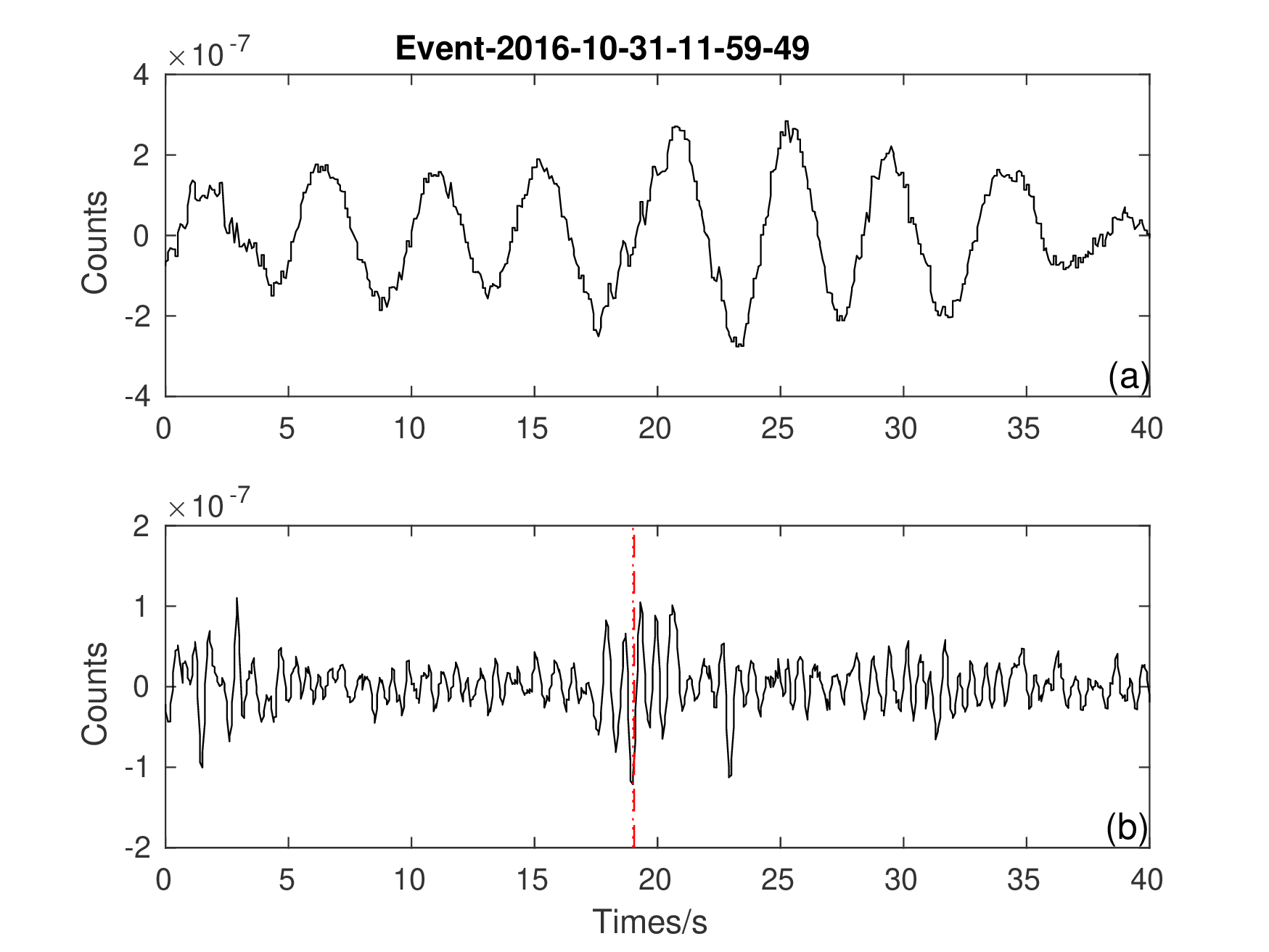} 
\caption{{\footnotesize (a) is the seismogram for the Event  609632174 (Table 1) in IASPEI  GT reference events list, only the part between 0.0s and 40s after P-wave arrival is shown. (b)  The series of reflectivity obtained by MED (Eq. (\ref{eq2})). The red dashed line and dotted line mark the times occupying the first smallest and the second smallest amplitude, respectively.} }\label{fig5}
\end{figure}

We have hidden more than two discrete unit impulses into the seismogram we cut, and all of them can be detected with the MED technique. They are also located at the times at which the amplitudes output from the MED are more smaller than those from the other phases or noises in the seismogram.

\begin{figure}
\includegraphics[scale=0.5]{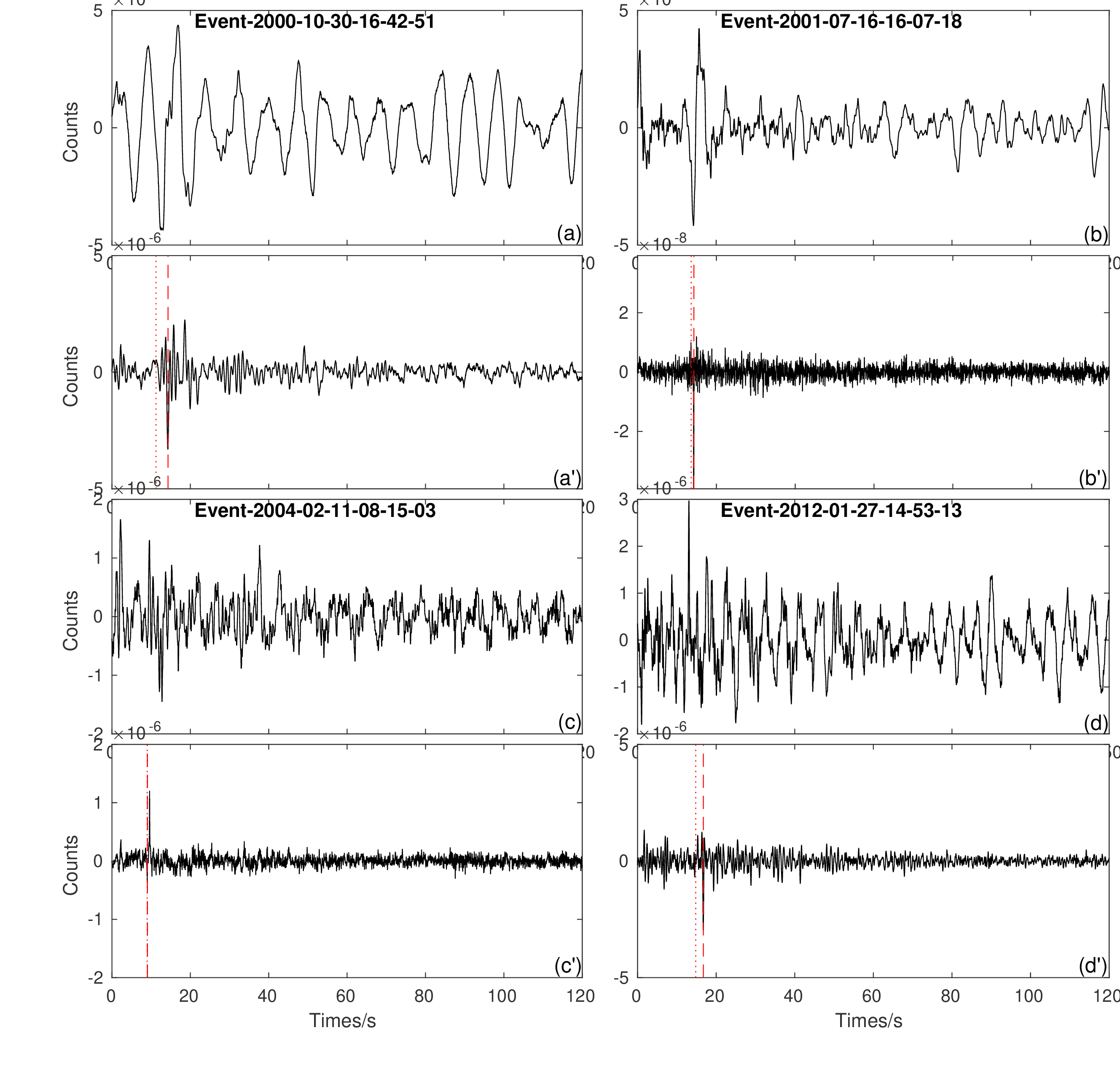} 
\caption{{\footnotesize (a), (b), (c), (d) are the seismograms for the Event 1741194, 3160315, 7250126, and 604758676 (Table 1), respectively. (a'), (b'), (c'), and (d') are the corresponding series of reflectivity obtained by MED (Eq. (\ref{eq2})), respectively. The red dashed line in each subfigure of reflectivity series (eg., (a')) marks the time occupying the smallest amplitude; The red dotted line marks the time difference pP-P from the IASPEI  GT bulletin.} }\label{fig6}
\end{figure}

Figure {\ref{fig3}} shows the result from burying a Gaussian mono-impulse into the seismogram we used.  The impulse has a starting amplitude of $-0.4A_{\rm{max}}$.  The series of reflectivity are in Figure {\ref{fig3}}. (d), and we also mark the true times at which the Gaussian mono-impulse starts and ends with red dashed lines in this subfigure. It can be found that these dashed lines are almost coincident with the lines occupying the smallest amplitudes but with a little difference. Further analysis shows the time with the smallest amplitude is close to the end time of the Gaussian mono-impulse, while the time with the second smallest amplitude is close to the start time of this impulse. These above show that the Gaussian mono-impulse buried can be detected with MED but it has a little delay.

Figure {\ref{fig4}} shows the result from hiding two Gaussian mono-impulses into the seismogram we used.  The impulses have a starting amplitude of $-0.4A_{\rm{max}}$ and $-0.6A_{\rm{max}}$, respectively. The series of reflectivity are in Figure {\ref{fig4}}. (d), and we also mark the true times at which the Gaussian mono-impulses start and end with red dashed lines in this subfigure. It can be found the same properties as those when a single Gaussian mono-impulse is added. Further analysis shows the time with the smallest amplitude is close to the end time of the first Gaussian mono-impulse, while the time with the second smallest amplitude is close to the start time of this impulse; The time with the third smallest amplitude is close to the end time of the second Gaussian mono-impulse, while the time with the fifth smallest amplitude is close to the start time of this impulse; The time for the fourth smallest amplitude is close to the time origin of the time window. These above show that two Gaussian mono-impulses buried can be detected with MED but they have a little delay. Similar properties can be found for the cases when more than two Gaussian mono-impulses are added into the seismogram here.

\section{Application to 12 Events in IASPEI  GT reference events list }

pP phase is a near-source solid-surface reflection, but otherwise follow approximately the same path as the principal $P$ phase through the rest of the earth. We assume that this reflective phase is an impulse-like signal in the $P$ coda, and it corresponds to the smallest amplitude of the series of reflectivity obtained by MED. An example for verification of the feasibility of this assumption can be found from Figure \ref{fig5}. Figure \ref{fig5} (a) is the seismogram for the Event  609632174 (Table \ref{Tb1}) in IASPEI GT reference events list (We cut the seismogram using between 0.0s and 120s but only show the part between 0.0s and 40s after P-wave arrival). Figure \ref{fig5} (b) plots the series of reflectivity obtained by MED (The data processing before MED will be explained below). The red dashed line and dotted line mark the times with the first smallest and the second smallest amplitude, respectively. It can be found that: (1) These two lines are very close with a time difference of 0.05s; (2) The time occupying the smallest amplitude is 19.05s, which is close to that for pP phase (17.4s after P arrival from the IASPEI  GT bulletin). Combining with the synthetic tests above, it is clear that pP phase can be described by a discrete unit impulses rather than a Gaussian mono-impulse (unless its duration is very very short), and it can be picked up through the smallest amplitude of the series of reflectivity obtained by MED. 
 
\begin{table}\scriptsize
\caption{{\footnotesize 12 earthquakes used from the IASPEI  GT reference events list}}\label{Tb1}
\begin{threeparttable}
\begin{tabular}{lllllll}
\hline
EVENTID\tnote{1}&Date and Time&Latitude($^\circ$)&Longitude($^\circ$)&Station& $t_{_{\rm{pP-P}}}$ (GT)\tnote{2} & $t_{_{\rm{pP-P}}}$ (MED)\tnote{3}\\  
\hline
1741194&2000/10/30/16:42:51&34.2920&136.2640&MVU&11.24s&14.30s\\
3160315&2001/07/16/16:07:18&32.8450&73.1300&DPC&13.6s&14.3s\\
7250126&2004/02/11/08:15:03&31.7106&35.4524&STU&8.99s&9.05s\\
604758676&2012/01/27/14:53:13&44.5433&10.0275&OBN&14.80s&16.75s\\
602754349&2013/03/31/07:02:37&42.6281&46.7839&KHC&10.90s&10.65s\\
602787145&2013/04/12/20:33:17&34.4134&134.8399&GUMO&1.34s&3.35s\\
605354327&2014/09/22/14:41:22&-40.5433&175.9272&FITZ&8.5s&9.0s\\
608619656&2016/04/11/19:41:06&-40.9108&175.5110&CTAO&5.83s&8.18s\\
611830427&2016/08/24/04:06:51&42.7612&13.1184&BILL&5.60s&5.35s\\
609439166&2016/09/11/13:10:08&42.0259&21.4598&WMQ&5.2s&6.4s\\
609632174&2016/10/31/11:59:50&45.8462&26.7773&ARU&17.40s&19.05s\\
610471076&2017/04/03/03:08:51&-26.8715&26.7661&DBIC&0.18s&0.05s\\
\hline
\end{tabular}
\begin{tablenotes}
        \footnotesize
        \item[1] Unique id number specific to each IASPEI reference event. 
        \item[2] Data for columns 1-6 is from International Seismological Centre (2019), IASPEI Reference Event (GT) List, https://doi.org/10.31905/32NSJF7V. 
        \item[3] Data is from this study.  
   \end{tablenotes}
    \end{threeparttable}
\end{table}

Based on the experiences above, we apply the MED technique to 12 earthquake events in the IASPEI  GT reference events list. Some information on these events is shown in Table \ref{Tb1}. As mentioned earlier, only the component of "BHZ" is used here for all the events. For each seismogram, data processing includes: (1) Removing the spikes with a median filter; (2) Interpolating the missing values; (3) Detrending the seismogram to remove the linear drift; (4) Cutting the seismogram using between 0.0s and 120s after P-wave arrival according to the IASPEI  GT bulletin; (5) Calculating the series of reflectivity obtained by MED; (6) Picking up the pP phases through the smallest amplitude of the reflectivity series in (5).   

\begin{figure}
\includegraphics[scale=0.5]{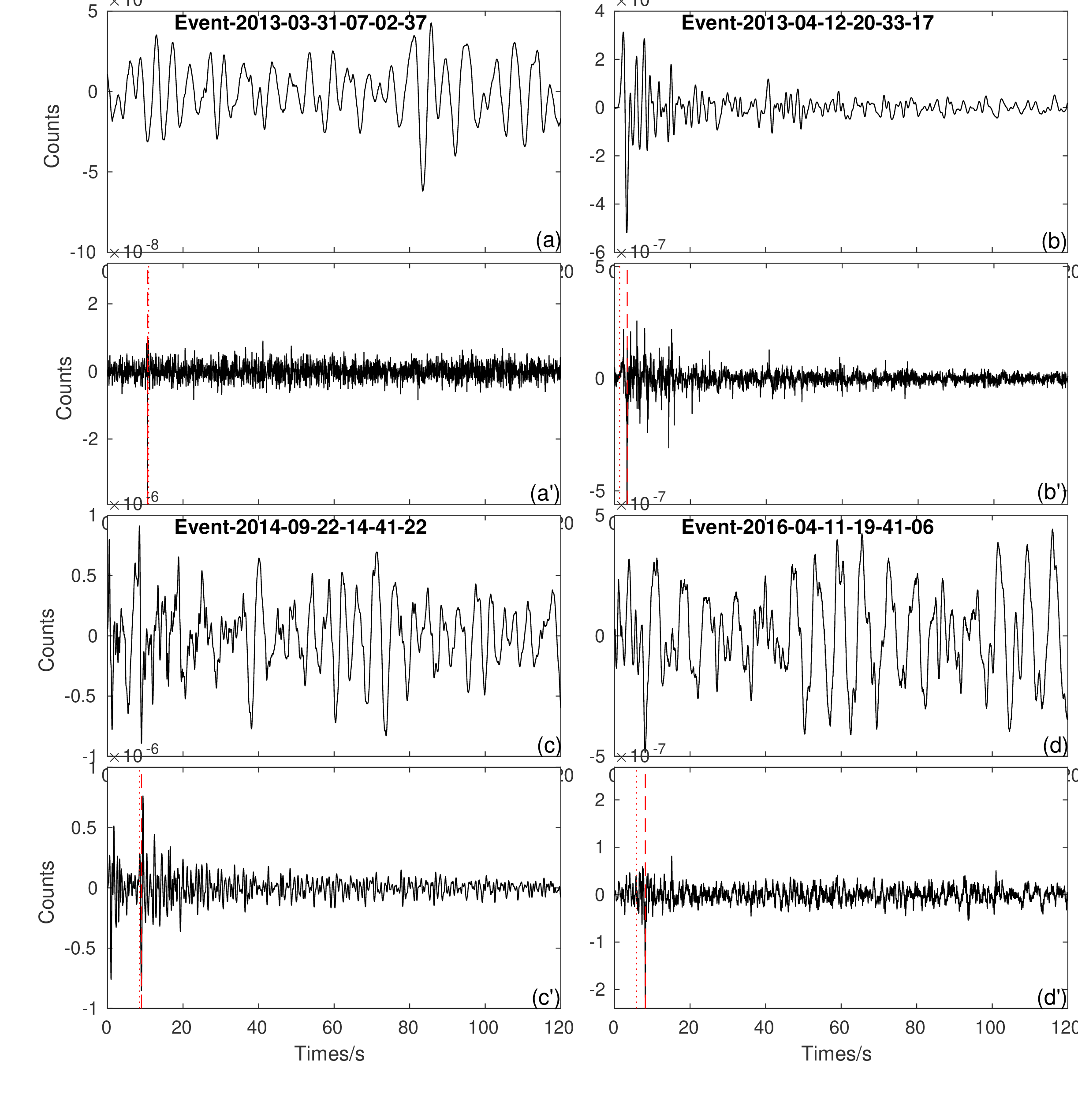} 
\caption{{\footnotesize(a), (b), (c), (d) are the seismograms for the Event 602754349, 602787145, 605354327 and 608619656 (Table 1), respectively. (a'), (b'), (c'), and (d') are the corresponding series of reflectivity obtained by MED (Eq. (\ref{eq2})), respectively. The red dashed line in each subfigure of reflectivity series (eg., (a')) marks the time occupying the smallest amplitude; The red dotted line marks the time difference pP-P from the IASPEI  GT bulletin.} }\label{fig7}
\end{figure}

\begin{figure}
\includegraphics[scale=0.5]{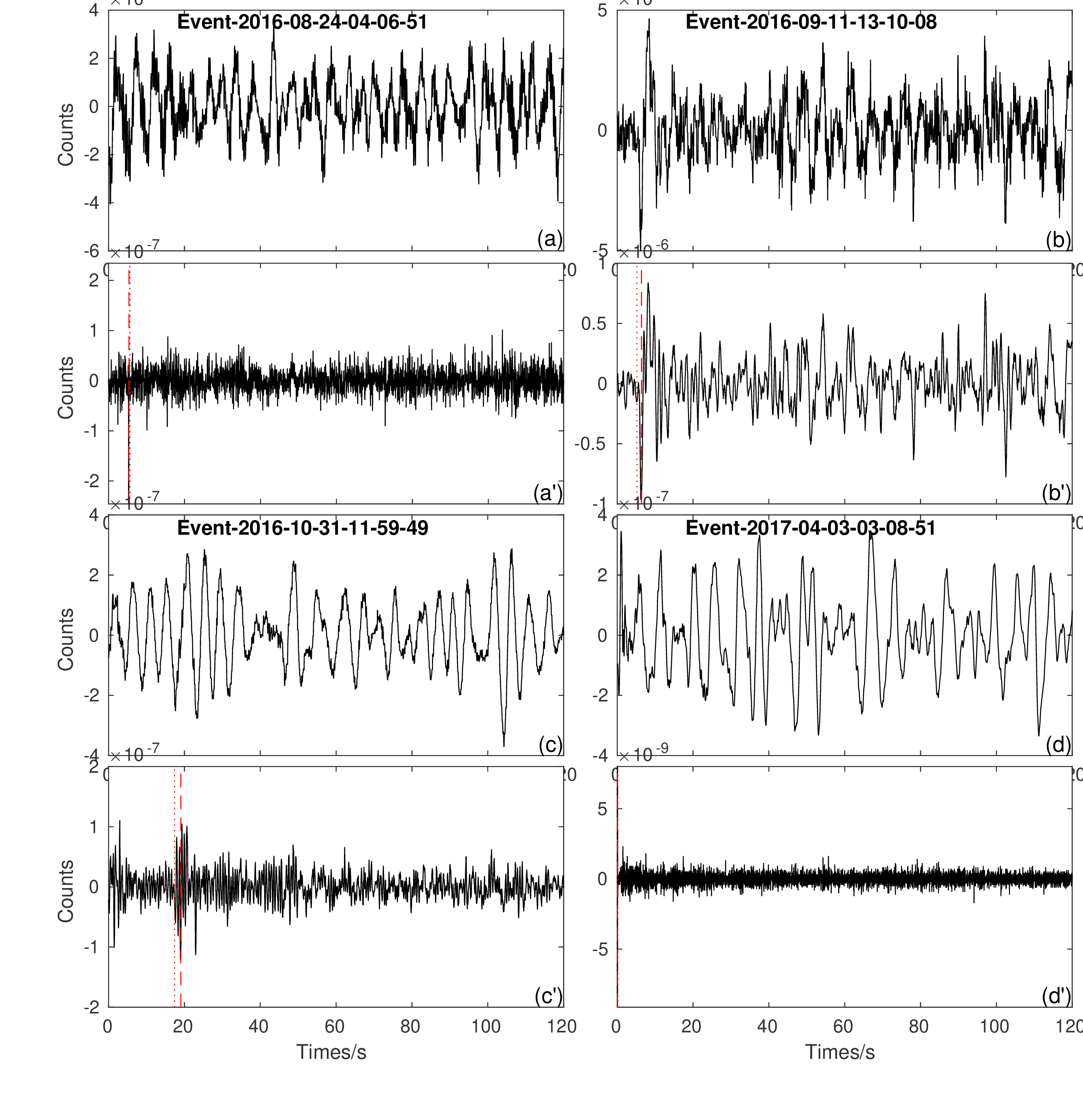} 
\caption{{\footnotesize (a), (b), (c), (d) are the seismograms for the Event 611830427, 609439166, 609632174, and 610471076 (Table 1), respectively. (a'), (b'), (c'), and (d') are the corresponding series of reflectivity obtained by MED (Eq. (\ref{eq2})), respectively. The red dashed line in each subfigure of reflectivity series (eg., (a')) marks the time occupying the smallest amplitude; The red dotted line marks the time difference pP-P from the IASPEI  GT bulletin.} }\label{fig8}
\end{figure}

Results are shown in Table \ref{Tb1} and Figure \ref{fig6}-\ref{fig8}. In each of the Figure \ref{fig6}-\ref{fig8}, (a), (b), (c), (d) are the seismograms for the 4 Events (Table 1), respectively. (a'), (b'), (c'), and (d') plot the corresponding series of reflectivity obtained by MED (Eq. (\ref{eq2})), respectively. The red dashed line in each subfigure of reflectivity series marks the time occupying the smallest amplitude, namely the time difference pP-P; The red dotted line marks  time difference pP-P from the IASPEI  GT bulletin. It can be seen from these figures that two lines (dashed and dotted line) in Figure \ref{fig6} (b'), Figure \ref{fig6} (c'), Figure \ref{fig7} (a'), Figure \ref{fig7} (c'), Figure \ref{fig8} (a'), Figure \ref{fig6} (d') are very close, while they are far apart in Figure \ref{fig6} (a'). From Table \ref{Tb1} it can be found that the errors for event  3160315, 7250126, 602754349, 605354327, 611830427, and 610471076 are less than 1s, while error for event 1741194 is 3.06s. The smallest error is 0.06s for event 7250126, which also can be seen from Figure \ref{fig6} (c'). These results show the technique of MED works well and effectively even for a single seismogram. 

\section{Discussions}
\subsection{Limitations with the MED}
The method of MED has many advantages, but it also has some limitations. The first limitation with the MED is how to determine the length ($M$) of the deconvolution filter $\bf{\bar{w}(t)}$. Our experiences show that too long or too short $M$ will lead to different but incorrect results, and the appropriate $M$ depends on the input data. Unfortunately, such appropriate filter length $M$ cannot be foreseen straightforwardly. A good way is trying filters of different lengths and examining the outputs to search this filter length. Obviously, this is an empirical approach. Here we select the filter length according to a minimum change in Kurtosis norm between filtering lengths, which is a good approximation but still not the perfect way. Besides, Eq. (\ref{eq4}) is highly nonlinear so that it can be solved iteratively. The iterative solution may not lead to a unique maximum value for $V_{\bar{r}}$, and it needs a good initial solution. In our tests, there are failure examples which present incorrect arrival time of pP. Therefore sometimes this method will be used with the aid of other methods, to obtain a reliable solution.

Here we assume that only pP phase is included in the P coda, it corresponds to the smallest amplitude of the reflectivity series from the MED. This is the second limitation, because there may be other solid-surface reflected waves whose energy is stronger than pP phase, perhaps sP phase for example. At this time, our assumption are likely to obtain sP phase rather than pP phase. Besides, we have not considered the polarity of pP waves. If the polarity of pP phase is opposite to our assumption, it should be picked up from the positive amplitude in the reflection series from the MED. But these problems arise not only for MED, but also for other technique. In this sense, MED may be useful in obtaining an independent starting point for other schemes which require a reasonable arrival of pP phase.

The third limitation is that we assume that pP phase is an impulse-like signal in the $P$ coda, while sometimes pP is more likely to be a wide harmonic. At this time, we may detect its end time instead of its start time with the MED technique. In this way, the time for pP phase obtained will be delayed.

\subsection{Another technique to detect echo based on MEC}

In many cases (for example in the cepstral method) a signal $x(t)$ is assumed as the sum of a direct wave $f(t)$ and one or two echoes which have the same frequency content as the $f(t)$ with only delays in time, as indicated in Eq. (\ref{eq5}):

\begin{equation}\label{eq5}
x(t)=f(t)+a_1f(t-t_1)+a_2f(t-t_2)
\end{equation}
where $a_1, a_2\in [-1,1]$ are two amplitude coefficients; $t_1,t_2$ are time delays.

\begin{figure}
\includegraphics[scale=0.5]{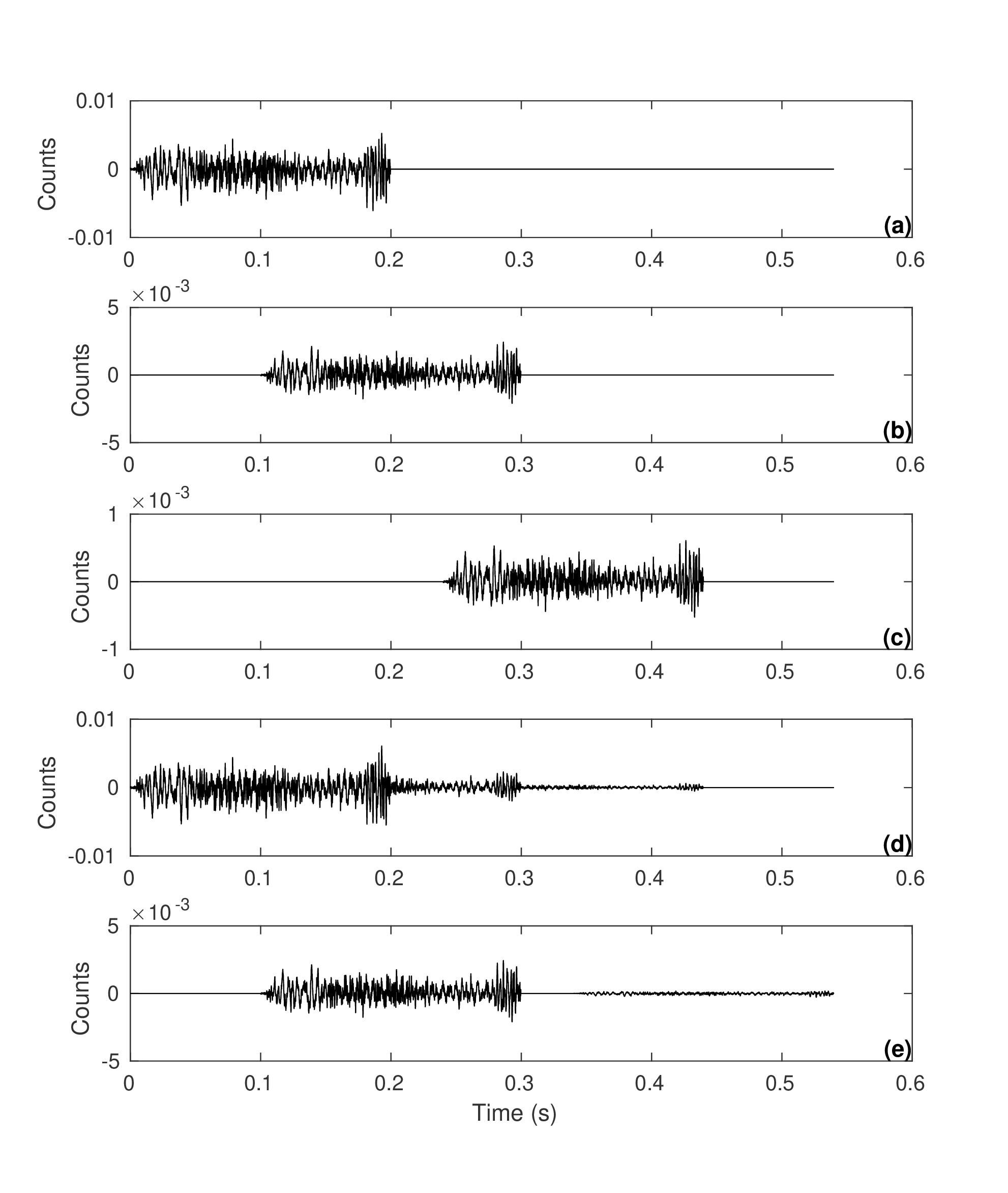} 
\caption{{\footnotesize (a) The direct wave $f(t)$.  (b) The first echo $-0.4f(t-0.1)$. (c) The second echo $-0.1f(t-0.24)$. (d) The signal $x(t)=f(t)-0.4f(t-0.1)-0.1f(t-0.24)$. (e) The echo $-0.4f(t-0.1)$ picked up with the technique of autocorrelation adding filtering based on MEC.}} \label{fig9}
\end{figure}

\begin{figure}
\includegraphics[scale=0.5]{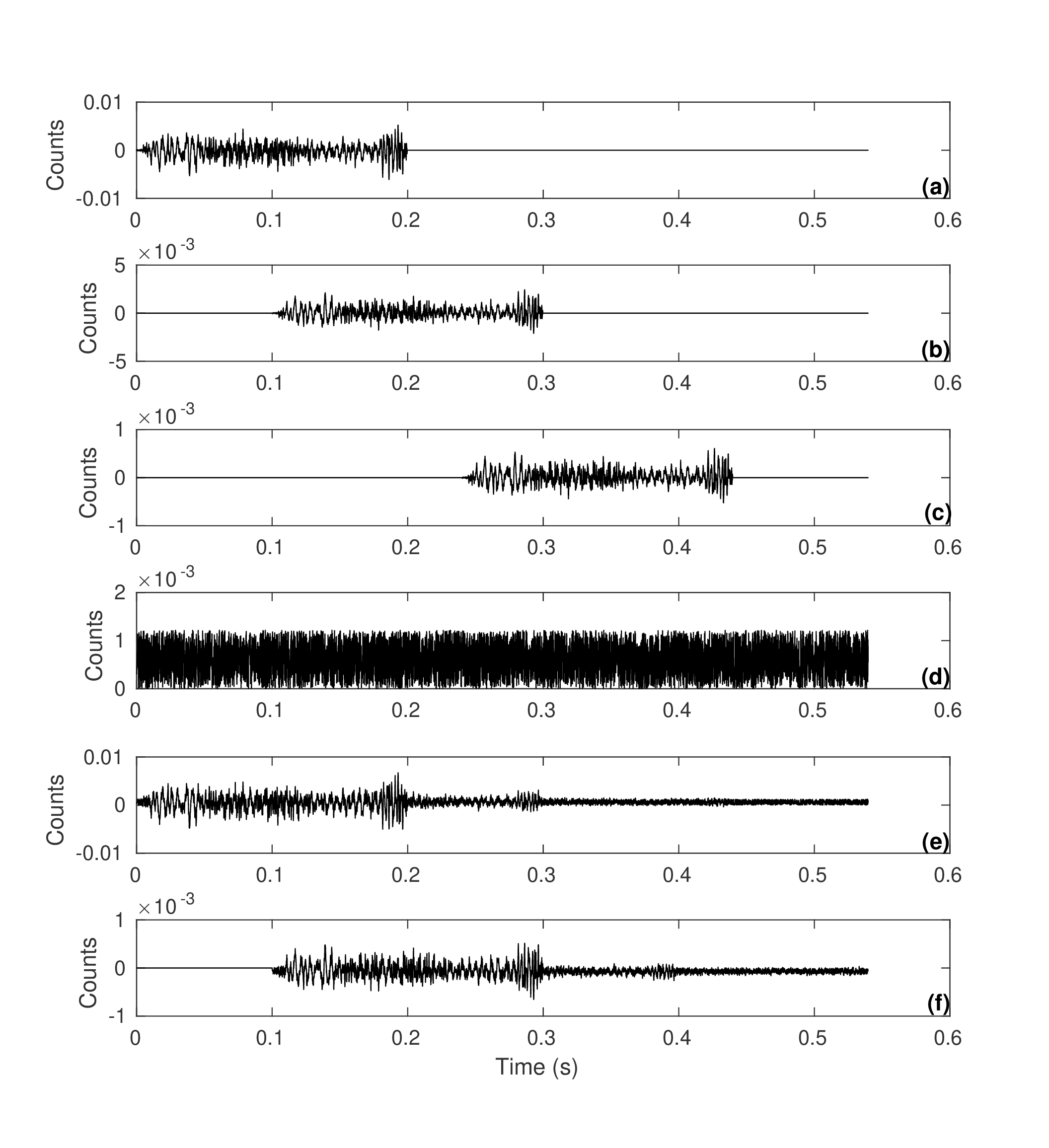} 
\caption{{\footnotesize (a) The direct wave $f(t)$.  (b) The first echo $-0.4f(t-0.1)$. (c) The second echo $-0.1f(t-0.24)$. (d) The random noises added to $x(t)$. (e) The signal $x(t)=f(t)-0.4f(t-0.1)-0.1f(t-0.24)+\rm{noises}$. (f) The echo $-0.4f(t-0.1)$ picked up with the technique of autocorrelation adding filtering based on MEC.}} \label{fig10}
\end{figure}

If $a_1 (\vert a_1\vert>\vert a_2\vert)$ or $a_2 (\vert a_1\vert < \vert a_2\vert)$, is known, $f(t-t_1)$ or $f(t-t_2)$ can be exactly determined with the technique of autocorrelation adding filtering (eg., Buck et al., 2002). However, in most cases,  $a_1$ or $a_2$ are unknown. With the MEC, ie., by maximizing the Kurtosis norm of the signal filtered, we can exactly pick up the $f(t-t_1)$ or $f(t-t_2)$. An example is shown in Figure \ref{fig9}, in which the $x(t)=f(t)-0.4f(t-0.1)-0.1f(t-0.24)$ and the echo of $0.4f(t-0.1)$ is to be picked up. Assuming that $a_1=-0.4$ is not known in advance, we use the technique of autocorrelation adding filtering based on MEC to detect it. Figure \ref{fig9} (e) demonstrate the echo of $0.4f(t-0.1)$ is detected successfully, even we add some random noises to $x(t)$ (Figure \ref{fig10} (f)). The delay time is picked up exactly and simultaneously.

The cases when there is only one echo, or there are three echoes are also tested with the technique of autocorrelation adding filtering based on MEC here. The echo interested (with larger absolute amplitude coefficient) is successfully detected, and the delay time is picked up exactly when there are no random noises simultaneously. When there are random noises to be added to $x(t)$, sometimes the echo interested is not detected, especially the amplitude of the noises is larger than that of $x(t)$. This is the same to the case when there are two echoes above.  

Unfortunately, we have not applied successfully this technique of autocorrelation adding filtering based on MEC to an actual seismogram. One possible reason is that the actual seismic waveforms we test do not meet the assumptions here. But this technique is worthy of further study. 

\subsection{$t_{_{\rm{pP-P}}}$ varies with epicentral distance}

\begin{figure}
\includegraphics[scale=0.5]{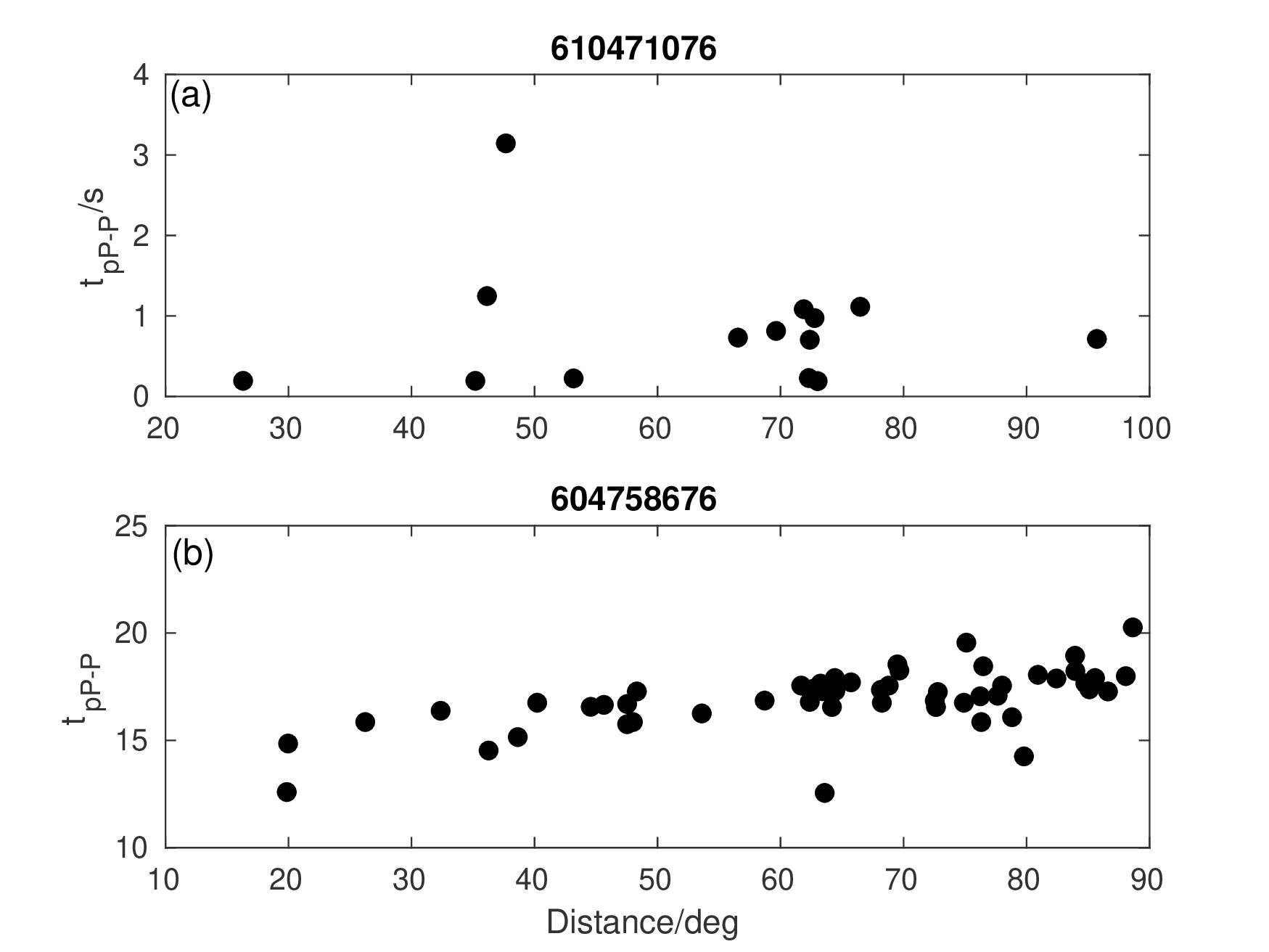} 
\caption{{\footnotesize Variation of the $t_{_{pP-P}}$ with epicentral distance for two events in the IASPEI GT reference events list.}} \label{fig11}
\end{figure}

As mentioned in the introduction, there is indeed a little moveout of the pP depth phases with increasing epicentral distance. That is, the travel-time differences of pP-P varies with epicentral distance. Craig (2019) demonstrate this with a synthetic example. Here we further illustrate this with the $t_{_{pP-P}}$s from the IASPEI  GT bulletin for two events (610471076 and 604758676, Table \ref{Tb1}) in the IASPEI GT reference events list, as shown in Figure \ref{fig11}.  It can be found that the moveout of the pP depth phases is clear, although there is scatter in the data. This should to be noted when we apply the usual stacking approaches to improve the detection of low-amplitude pP phases.  Although the methods based on stacking may be more and more common in the future, this moveout of the pP demonstrate further that it is still necessary to develop new techniques to detect pP phase from the seismograms observed only on a single-station.

\section{Conclusions}

The technique of MED is effective to detect the pP-like phase when the seismic data meets the corresponding mathematical models and assumptions. Such effectiveness has been demonstrated from synthetic waveforms applications. MED is also effective to pick up pP phase from the actual seismogram, and the key to MED is how to determined the length of the deconvolution filter. It is a good way to select this filtering length according to a minimum change in Kurtosis norm between filtering lengths from our experiences.    

\vspace{5em}


\ \ 

\end{document}